\documentclass[aps,prl,showpacs,floatfix,reprint,superscriptaddress]{revtex4-1}
\usepackage{etoolbox}  
\usepackage{graphicx} 
\usepackage{hyperref}
\usepackage{amsmath}
\usepackage{amssymb}
\usepackage{ifsym}
\usepackage{epstopdf}
\usepackage{siunitx}
\usepackage{lipsum}
\usepackage{amssymb,amsthm,amsmath,hyperref,epsfig,wrapfig}

\newcommand{\BaCo}{Ba(Fe$_{1-x}$Co$_x$)$_2$As$_2$}

\newcommand{\Rff}{$\rho_{\rm ff}$}
\newcommand{\jc}{$j_{c}$}

\begin{document}

\title{Relationship between critical current and flux-flow resistivity in the mixed state of \BaCo}

\author{X. Y. Huang}
\affiliation{Department of Physics, Kent State University, Kent, Ohio, 44242, USA}

\author{Y. P. Singh}
\altaffiliation[Present Address: ]{Department of Mechanical Engineering, The University of Akron, Akron, Ohio, 44325, USA}
\affiliation{Department of Physics, Kent State University, Kent, Ohio, 44242, USA}

\author{D. J. Haney}
\affiliation{Department of Physics, Kent State University, Kent, Ohio, 44242, USA}

\author{T. Hu}
\affiliation{Department of Physics, Kent State University, Kent, Ohio, 44242, USA}
\affiliation{Shanghai Institute of Microsystem and Information Technology, Shanghai 200050, China}

\author{H. Xiao}
\affiliation{Department of Physics, Kent State University, Kent, Ohio, 44242, USA}
\affiliation{Center for High Pressure Science and Technology Advanced Research, Beijing 100094, China}

\author{Hai-Hu Wen}
\affiliation{Nanjing University, Nanjing 210093, China}

\author{Shuai Zhang}
\altaffiliation[Present Address: ]{Beijing National Laboratory for Condensed Matter Physics and Institute of Physics, Chinese Academy of Sciences, Beijing 100190, China}
\affiliation{Department of Physics, Kent State University, Kent, Ohio, 44242, USA}

\author{M. Dzero}
\affiliation{Department of Physics, Kent State University, Kent, Ohio, 44242, USA}

\author{C. C. Almasan}
\affiliation{Department of Physics, Kent State University, Kent, Ohio, 44242, USA}

\begin{abstract}
We studied the temperature and magnetic field dependence of  vortex dissipation and critical current in the mixed-state of unconventional superconducting alloys \BaCo ~($0.044 \leq x \leq 0.100$) through current-voltage measurements. Our results reveal that all the electric field $E$ vs current density $j$ curves in the Ohmic regime merge to one point ($j_0,E_0$) and that there is a simple relationship between the critical current density $j_c$ and flux-flow resistivity \Rff: $\rho_{\rm ff}/\rho_{\rm n} = (1- j_{c}/j_{0})^{-1}$, where $\rho_{\rm n}=E_0/j_0$ is the normal-state resistivity just above the superconducting transition. In addition, $E_0$ is positive for all five dopings, reflecting the abnormal behavior of the flux-flow resistivity \Rff: it increases with decreasing magnetic field. In contrast, $E_0$ is negative for the conventional superconductor Nb since, as expected, \Rff ~decreases with decreasing magnetic field. Furthermore, in the under-doped and over-doped single crystals of \BaCo, the parameter $E_0$ remains temperature independent, while it decreases with increasing temperature for the single crystals around optimal doping ($ 0.060\leq x\leq 0.072 $). This result points to the co-existence of superconductivity with some other phase around optimal doping. 
\end{abstract}

\maketitle

\section{Introduction}

The cobalt-doped superconducting iron-arsenide material \BaCo~ has been widely studied in part due to its rich phase diagram. Of special interest is the region of the temperature-doping ($T$-$x$) phase diagram in which the spin-density-wave (SDW) phase  \cite{FernandesPRB2010,RotunduPRB2011Secondorder,Chu13082010,Chu10082012,FernandesPRB2010} is in macroscopic coexistence with the superconducting (SC) phase \cite{CurroPRL2013}. With an increase in cobalt concentration, the SDW order is suppressed providing the stage for the possible existence of a quantum phase transition under the superconducting dome \cite{FanlongNingJPSJ2009,fernandesPRL2013howmanyQCP,ThermPowerQCP}. In addition, neutron scattering and nuclear magnetic resonance studies of BaFe$_2$(As$_{1-x}$P$_x $)$_2$ as a function of isovalent phosphorous doping have shown that the second-order SDW phase transition present at low phosphorous doping changes into a weakly first order transition in the vicinity of the optimally-doped sample  \cite{HuDingPRL2015}. These experimental facts have been used to  propose a scenario in which quenched order gives rise to a spatially inhomogeneous emulsion with puddles of SC and SDW phases \cite{ChowdhuryPRB2015}.   

Critical current density ($j_c$) and flux-flow resistivity (\Rff) have each been widely studied both in conventional type-II superconductors \cite{D.Hughes1974,Y.B.Kim1964,Y.B.Kim1965} and in unconventional superconductors such as iron-pnictide systems \cite{ProzorovPRB2008,TanatarIOP2010,J.HanischIEEE2011,MaedaPRB2012,MaedaPhyC2013,MaedaPhyC2014,MaedaPRB2015}. Nevertheless, to our knowledge there has not been any study revealing and discussing the presence of a relationship between these physical quantities. To address this issue, we carried out current-voltage ($I$-$V$) measurements on five superconducting \BaCo~single crystals with Co doping $x$ within the range $0.044 \leq x \leq 0.100$. We have discovered the following linear relationship between $j_c$ and inverse of flux-flow resistivity \Rff$^{-1}$: 
\begin{equation}\label{Eq1}
\rho_{\rm ff}^{-1}= \frac{j_0}{E_0}\cdot\left(1-\frac{j_c}{j_0}\right),
\end{equation}
where $E_0>0$ for all five dopings. Moreover, the analysis of our data shows that $E_0/j_0$ is the normal-state resistivity. We further show that such a relationship also exists in conventional type-II superconductors such as niobium, in which $E_0$, in contrast with \BaCo, has a negative value. In addition, the value of $E_0$ is temperature independent in under-doped and over-doped single crystals of \BaCo, whereas it decreases with increasing temperature for samples around the optimal doping ($ 0.060\leq x\leq 0.072 $). This latter experimental fact
is consistent with the existence of a secondary phase such as SDW, glass phase, or recently proposed micro-emulsion phase present in this region of the phase diagram. 

\section{Experimental details}

Single crystals of \BaCo~were grown using the FeAs self-flux method \cite{ChuPRB2009,NiNiPRB2008EffectofCo}. Their actual Co-doping value $x$ was determined by comparing the values of their superconducting critical temperature $T_{c0}$ at zero magnetic field with values from well-established $T_{c0}$-$x$ phase diagrams \cite{NiNiPRB2008EffectofCo,ChuPRB2009,J.Reid_PRB2010}. The Co-doping values of the single crystals discussed in this paper are $x=$0.044, 0.056, 0.060, 0.072, and 0.100, which cover the under-doped, optimally-doped, and over-doped regions. The temperature ($T$) and magnetic field ($H$) dependences of \Rff~were obtained from $I$-$V$ measurements carried out on thin samples using the standard four-probe method with current flowing in the $ab$-plane and magnetic fields applied along the crystallographic $c$-axis. Due to the high current required to depin the strongly pinned flux vortices in the mixed state of these superconductors, a combination of Linear Research, Inc. LR700 resistance bridge with extended current limit and Physical Property Measuring System (PPMS) were used to carry out the $I$-$V$ measurements. 

We reduced Joule heating of the single crystals due to the high current used in the $I$-$V$ measurements as follows: (1) the cross-section area of the single crystals was reduced down to $0.17\times0.04$ mm$^2$ in order to increase the actual current density. The reason is that, for a given heating power per unit length $p\equiv(I^2R)/L=(j^2A^2)\cdot(\rho L/A)/L= j^2\rho A$, a maximum current density $j$ is accomplished for an achievable minimum cross-section area $A$; (2) multiple short thick gold current leads were used for the two current terminals to minimize the Joule heating in the gold wires  and also to increase the heat transport from the single crystal to the thermal bath, since, for a given applied current, the dissipated power $ P= I^2R = I^2\rho L/A$; (3) we used Sn, instead of silver paste, in order to decrease the contact resistance between the single crystal and the current leads down to less than 10 $\mu\Omega$ \cite{TanatarIOP2010}; (4) an additional temperature sensor mounted on the top of the sample using $N$-type grease was used to control and measure the temperature of the sample and long folded manganin wires were used as the terminal leads of the thermometer to decrease the heat transport between the thermometer and puck since manganin has poor thermal conduction; (5) after the temperature was deemed stable, a 60 sec wait time (with the persistent current flowing through the single crystal) was included into the measurement sequence and only then the $I$-$V$ data were collected. 

The upper limit of the current used in our measurements was sample dependent. We determined this value experimentally by monitoring the temperature of the single crystal with the temperature sensor mounted on the crysta,l as discussed above, and by stopping the current sweep when this temperature increased by more than 0.1 K from the desired temperature, an increase due to Joule heating. Hence, with all the improvements discussed above, we were able to get reliable  $I$-$V$ data for these \BaCo~single crystals by reducing the temperature instability due to Joule heating in the mixed state to less than 0.1 K. In addition, due to the high current limit imposed by our experimental condition, all the $I$-$V$ measurements were done at temperatures 0.85 $T_{c0}\leq T\leq$ 0.98 $T_{c0}$, i.e. about 2 K below the $H$-$T$ phase boundary. 

\begin{figure}
\centering

\includegraphics[width=1\linewidth]{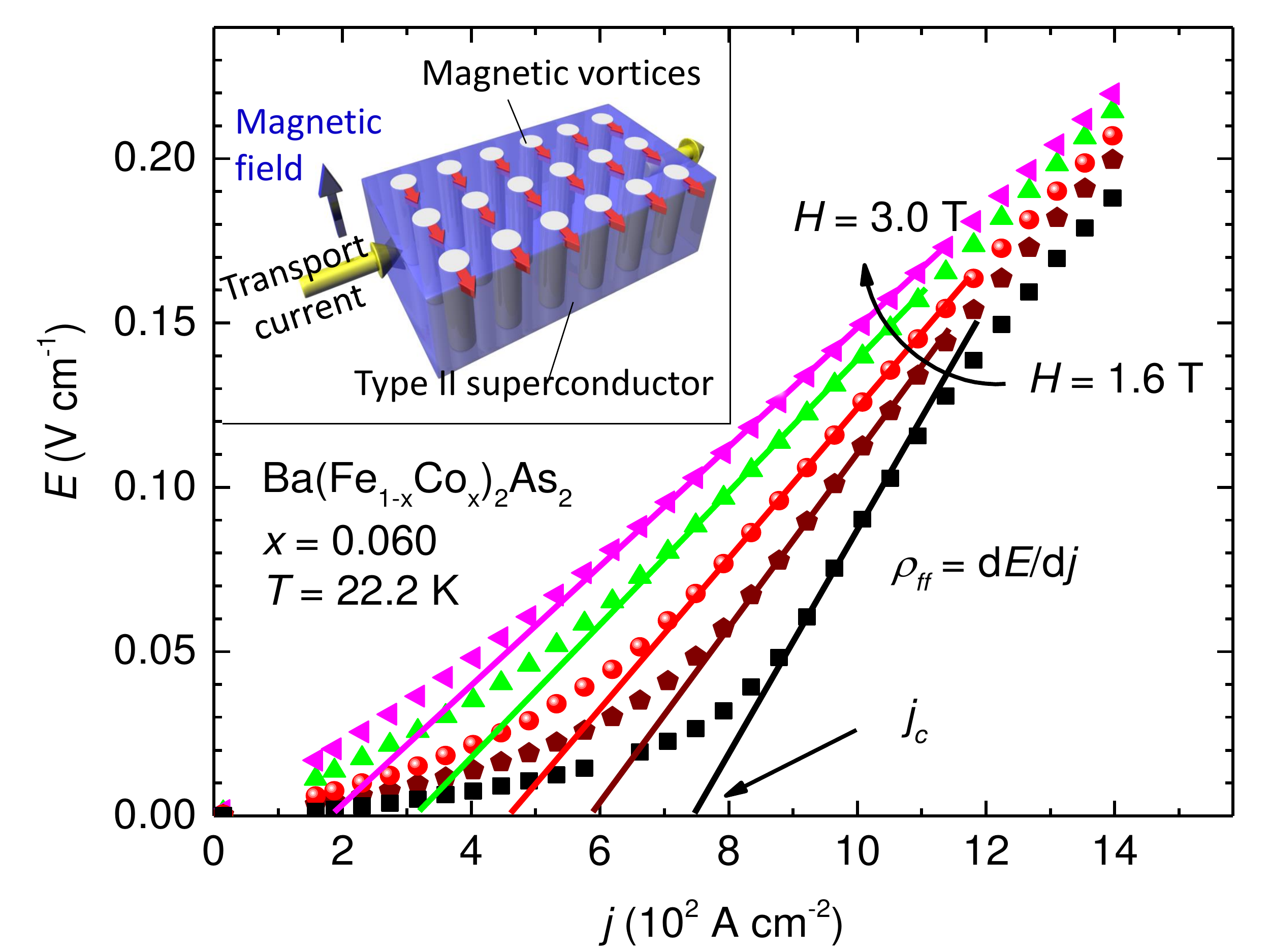}
\caption{\label{f1}Electric field - current density $E$-$j$ characteristics for the \BaCo~single crystal with $x = 0.060$ measured at a temperature $T = 22.2 $ $\si{\kelvin}$ and for applied magnetic fields $H = 1.6, 1.8, 2.0, 2.4, 3.0$ T. Inset: Schematic illustration of a type-II superconductor (blue rectangular parallelepiped) and the flux vortices (tubes) that exist when this superconductor is in the mixed state and in the presence of an applied magnetic field perpendicular to the applied current.}
\end{figure}

\section{Results and discussion}

\subsection{Electric field - current density curves below $T_c$} 
We show in inset to Fig. \ref{f1} schematically a type-II superconductor in an external magnetic field: when the magnetic field is applied along the $c$ crystallographic direction (smallest dimension in the figure) and the applied current $I$ is along the $a$ crystallographic axis, the flux vortices are driven in the $b$ crystallographic direction resulting in a longitudinal dissipative voltage along the current direction.  An $I$-$V$ curve is generated by increasing the applied current and measuring the longitudinal  voltage. 

We plot in Fig.~\ref{f1} the electric field  - current density ($E$-$j$) curves obtained, by taking into account the sample geometry, from the $I$-$V$ characteristics at $T = 22.2$ K under fields of $1.6 \leq H \leq 3.0$ T for the \BaCo~single crystal with $ x= 0.060$. These data are typical for all the single crystals with different Co doping that we have measured. A voltage is first detected when vortices start to creep. By increasing the applied current, the flux vortices are driven harder and harder. A linear (Ohmic) regime in the $E$-$j$ curve appears when the vortices are fully de-pinned,  which corresponds to the flux-flow regime. The slope of the linear regime in the $E$-$j$ curve represents the flux-flow resistivity, i.e., \Rff~$=  dE/dj$. This physical quantity is dominated by the scattering of the quasiparticles in/around the vortex cores. The intercept of this linear regime with the $j$ axis gives the critical-current density $j_c$. With increasing $H$, the Ohmic regime increases and it covers the whole current range for $H\geq H_{c2}$ at the measured temperature. 

\subsection{Relationship between $j_{c}$ and \Rff\ }

\begin{figure}
\centering
\includegraphics[width=1.0\linewidth]{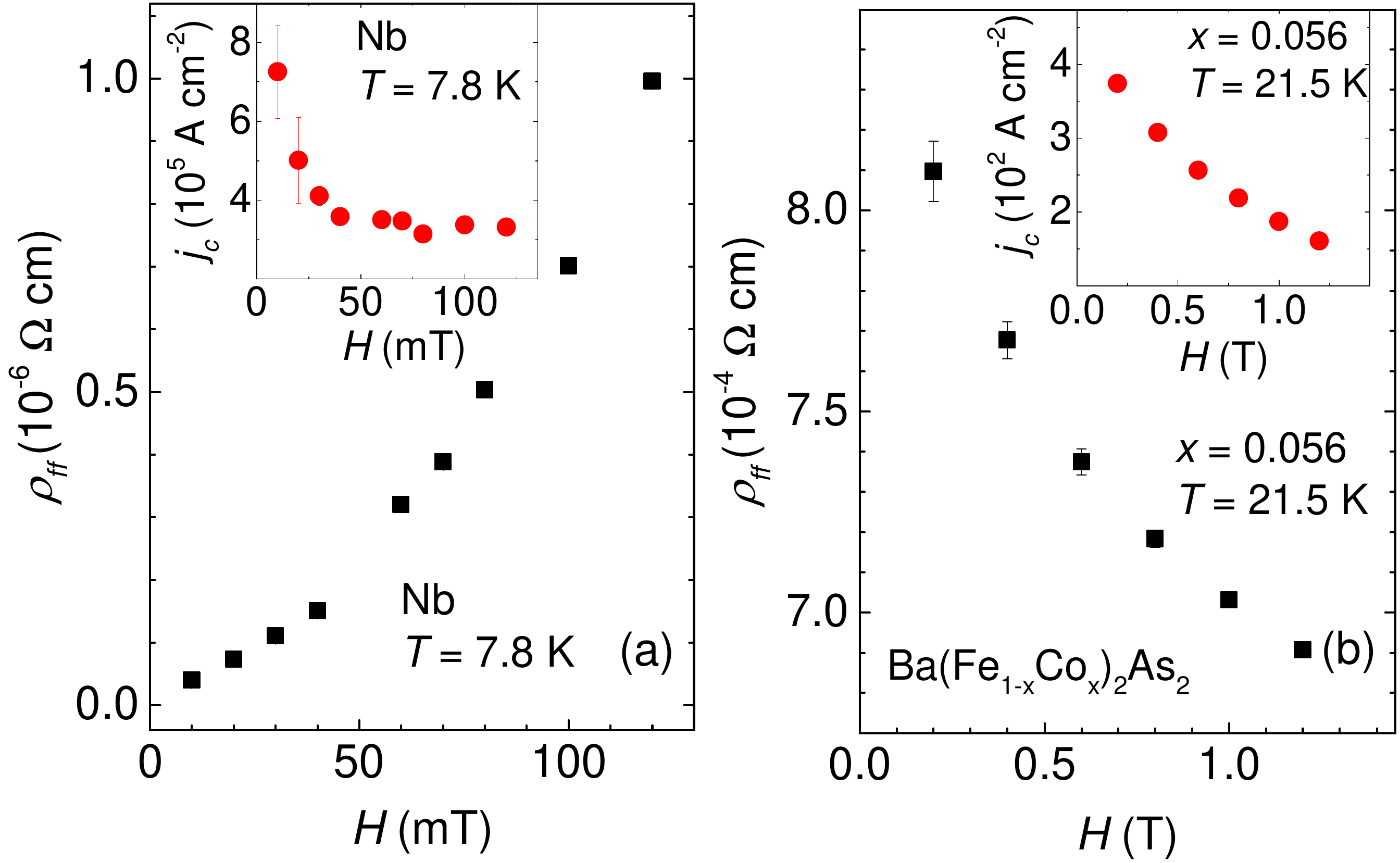}
\caption{\label{f2} 
(a) Magnetic field $H$ dependence of flux-flow resistivity  \Rff~(main panel) and critical current density $j_c$ (inset) of niobium (Nb) measured at a reduced temperature $t\approx 0.9$. Data are taken from Ref. \cite{C.PerozPRB2005}. (b)  $H$ dependence of \Rff~(main panel) and $j_c$ (inset) for \BaCo~with $x=0.056$, measured at $T = 21.5 $ $\si{\kelvin}$ ($t\approx0.9$) and for $0.2 $ T $ \leq $ $H \leq 1.2 $ T, namely, reduced magnetic field $ 0.05\leq h \leq 0.29$. We note that the measured reduced temperature and reduced magnetic field ranges are the same for Nb and \BaCo~systems.}. 
\end{figure}

It is well known that the flux-flow resistivity in the mixed state of conventional type-II superconductors is proportional with the magnetic field in the low-field and low-temperature regimes, i.e., there is the following empirical relationship \cite{Y.B.Kim1965}:
\begin{equation}
\frac{\rho_{\rm ff}}{\rho_{\rm n}} \propto \frac{H}{H_{c2}},
\end{equation} 
where $\rho_n \equiv \rho_{\rm ff}(H_{c2})$ \cite{BardeenStephen1965}, and it saturates near $H_{c2}$  \cite{KopninPRB1995,KoppinPRL1997,KambePRL1999UPt3}. Indeed, the flux-flow resistivity of, for example, niobium increases with increasing $H$ \cite{Huebener1970, C.PerozPRB2005}:  
Fig.~\ref{f2}(a) shows the increase of flux-flow resistivity with increasing $H$ up to 126 mT ($h=H/H_{c2}=0.3$, $H_{c2}=420~m$T), measured at a high reduced temperature ($t=T/T_{c0}\sim 0.9$) \cite{C.PerozPRB2005}.

Now we turn our attention to the experimental results obtained on the \BaCo~single crystals. 
We show \Rff($H$) measured on  \BaCo~with $x = 0.056$ at $T=21.5$ K and $0.2 $ T $ \leq $ $H \leq 1.2 $ T in Fig.~\ref{f2}(b). These measurements are done at a reduced temperature $t\approx 0.9$ and reduced field $ 0.05\leq h \leq 0.29$ since $T_{c0}=23.2$ K and $H_{c2}=4.2$ T at this temperature. Therefore, the data shown for Nb in Fig.~\ref{f2}(a) and  for \BaCo~with $x = 0.056$ in Fig.~\ref{f2}(b) were collected at the same reduced magnetic field and temperature. Nevertheless, notice that, in contrast with Nb, the flux-flow resistivity of \BaCo~decreases with increasing magnetic field. This behavior is typical for all the samples we have measured.  Hence,  these two systems show opposite field dependences of flux-flow resistivity.  

We note that a similar abnormal field dependence of flux-flow resistivity was observed in unconventional heavy-fermion superconductor CeCoIn$_5$ and it was shown to be the result of critical spin fluctuations \cite{TaoPRL}.  We have discussed the origin of the abnormal \Rff$(H)$~and its doping dependence in the \BaCo~system elsewhere \cite{Huang2016}. 

In contrast to different \Rff($H$) measured in Nb on one hand and \BaCo~ and CeCoIn$_5$ on the other hand, the insets to Figs.~\ref{f2} show similar  magnetic field dependences of the critical current densities of  Nb and \BaCo: $j_c$ decreases monotonically with increasing $H$. In fact, such $j_c(H)$ behavior is typical of type-II superconductors \cite{SummersIEEE1991, AttanasioPhysicaC1995,TraitoPRB2006,JensNature2015}: the critical current density required to depin the flux vortices decreases with increasing $H$. 

\begin{figure}
\centering
\includegraphics[width=1.0\linewidth]{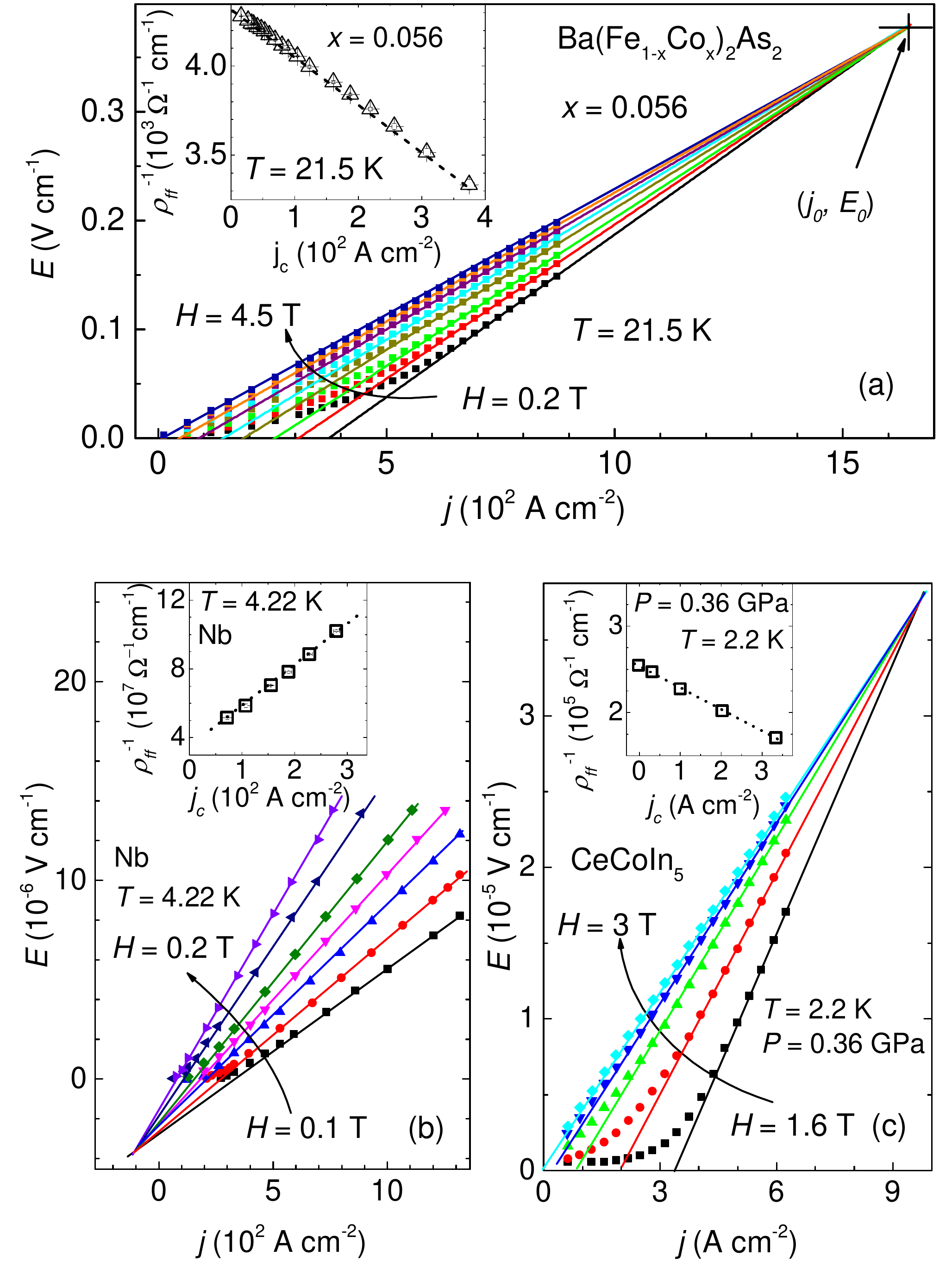}
\caption{ \label{f3}  Electric field - current density $E$-$j$ characteristics for (a) \BaCo~single crystal with $x = 0.056$ and $H=$ 0.2, 0.4, 0.6, 1.0, 1.4, 2.0, 3.0, 4.5 T, (b) Nb with data extracted from Ref. \cite{Huebener1970} and $H=$ 0.1, 0.1125, 0.125, 0.1375, 0.15, 0.175, 0.2 T, and (c) CeCoIn$_5$ with data extracted from Ref. \cite{TaoPRL} and $H=$  1.6, 1.8, 2.0, 2.2, 3.0 T. The marked point ($j_0$,$E_0$) in all these plots is the merging point of the Ohmic (linear) behavior of $E(j)$ in the flux-flow region. Insets: Corresponding inverse flux-flow resistivity \Rff$^{-1}$ vs  critical current density $j_c$.}
\end{figure}

In general, one would not expect a relationship between \Rff~and \jc~to exist since the former represents the dissipation in the free-flux-flow regime due to the quasiparticles present in/around the vortex cores and is independent of the pinning strength, while the latter reveals the strength of the pinning centers. However, plots of \Rff$^{-1}$~vs \jc~shown in the insets to Figs.~\ref{f3} reveal the presence of a linear relationship between these two quantities in \BaCo, Nb, and CeCoIn$_5$ systems, with a positive slope for the Nb sample and negative slopes for the \BaCo~ and CeCoIn$_5$ samples. 

We show in the main panels of Figs.~\ref{f3} the $E$ vs $j$ data for these three systems. For the \BaCo~ system we show the data for the $ x= 0.056$ single crystal as an example (the other Co-doped samples studied display a similar behavior). The \textit{linear} flux-flow regime is given by $E=\rho_{\rm ff}\cdot (j-j_{c})$. In addition, notice that all the fitting lines in the flux-flow regime merge at the same point ($j_0,E_0$) for all three systems. [The data do not extend all the way to this merging point ($j_0, E_0$) since we performed all the $E$-$j$ measurements at $H<H_{c2}$.] This implies that there is the following relationship between \Rff~and \jc~: 
\begin{equation}\label{Eq2}
\rho_{\rm ff}=\frac{E_0}{j_0-j_c},
\end{equation}  
from which Eq. (\ref{Eq1}) directly follows.
Therefore, the linear relationship between \Rff$^{-1}$~and \jc, shown by the insets of Figs.~\ref{f3}, is a result of the fact that the Ohmic regimes measured at different $H$ values merge in one point, denoted here ($j_0,E_0$). The merging point has a positive coordinate ($j_0,E_0$) for the \BaCo~ and CeCoIn$_5$ samples, and a negative coordinate for the Nb sample. This result is a consequence of the fact that \Rff$(H)$ decreases with increasing $H$ in \BaCo~ and CeCoIn$_5$, while it increases with increasing $H$ in Nb.

\begin{figure}
\centering
\includegraphics[width=1.0\linewidth]{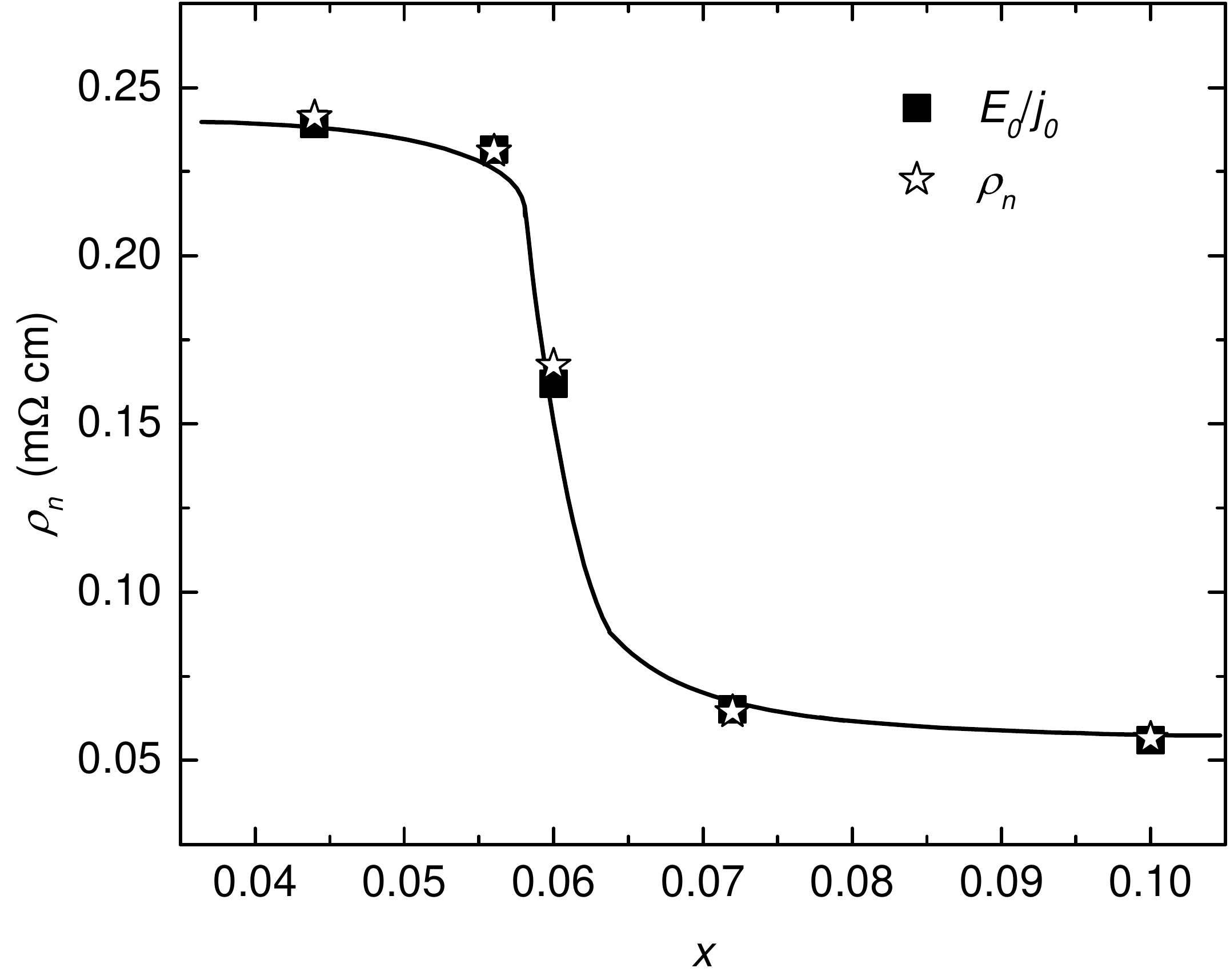}
\caption{\label{f4}Plots of experimentally-obtained normal-state resistivity $\rho_n$ at the superconducting transition and $E_0/j_0$  as a function of doping $x$ of the \BaCo~ system. The line is a guide to the eye.}

\end{figure}

The merging point ($j_0,E_0$) gives the resistivity in the normal state, i.e., $E_0/j_0 \approx\rho_n$. Indeed, Eq. (\ref{Eq1}) gives 
$\rho_{\rm ff}=E_0/j_0\approx \rho_n$ at $H=H_{c2}$ of a certain temperature since here $j_c=0$ and $\rho_{\rm ff}\approx \rho_n$ [see Eq. (2)].  We also plot $E_0/j_0$ vs $x$ along with the experimentally-obtained $\rho_n$ vs $x$ in Fig.~\ref{f4}. Notice the excellent agreement between these quantities at the same value of $x$. Also notice the change in the value of $\rho_n$ around optimal doping.
With regards to Nb, the value of $E_0/j_0=2.9\times 10^{-8}$ $\Omega \textrm{cm}$, obtained from Fig. \ref{f3}(b), is also in excellent agreement with  the experimentally obtained $\rho_{n}=2.66\times 10^{-8}$ $\Omega\cdot\textrm{cm}$ \cite{Huebener1970}. 

Based on the experimental fact that $E_0/j_0 \approx\rho_n$, we conclude that the Ohmic $E$-$j$ regimes measured at different $H$ values merge at the same point ($j_0, E_0$) in systems with very small normal-state magnetoresistivity. Therefore, the presence of the linear relationship between \Rff$^{-1}$~and \jc, given by  Eq. (1) and insets to Figs.~\ref{f3}, is the consequence of small magnetoresistance near $T_c$ in the systems discussed here. 

\begin{figure}
\centering
\includegraphics[width=1.0\linewidth]{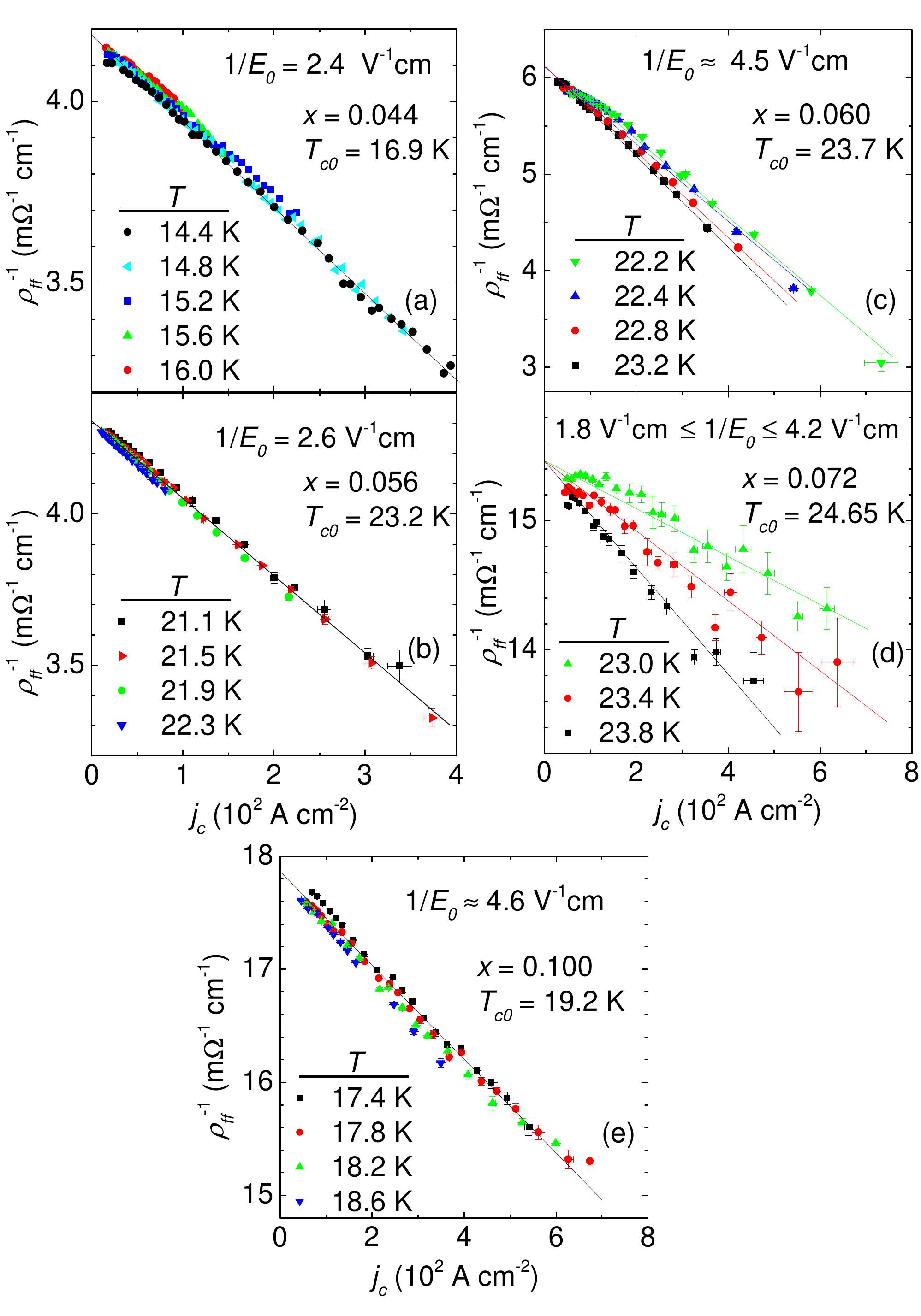}
\caption{\label{f5} Linear relation of $\rho_{\rm ff}^{-1}$ and $ j_{c} $ for five dopings of \BaCo~ system at multiple temperature below $T_c$. The temperature independence of slope of this relation is apparently shown in under-doped and over-doped ranges. Near optimally doped \BaCo, especially $x=0.072$, the slope,$1/E_{0}$, is sensitive to temperature.}

\end{figure}

\subsection{Relationship between \Rff~and $j_{c}$ at different temperatures}

\begin{figure}
\centering
\includegraphics[width=1.0\linewidth]{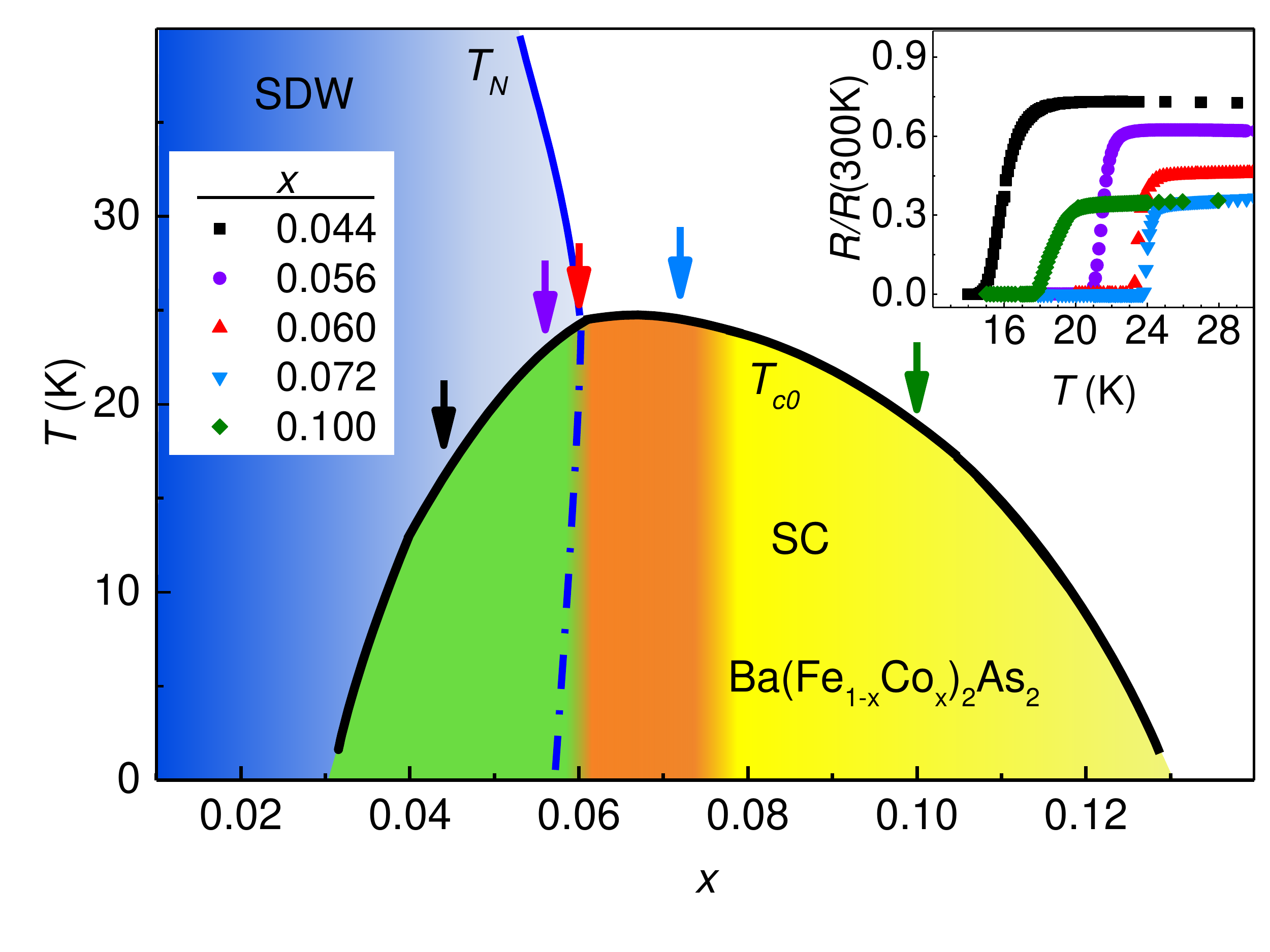}
\caption{\label{f6}$T$-$x$ phase diagram for  \BaCo~based on Refs. \cite{NiNiPRB2008EffectofCo,ChuPRB2009,FernandesPRB2010}. The arrows mark the five doping levels in our measurement; green, red, and yellow colors label the three ranges of doping in our discussion. The inset shows the normalized resistance vs temperature $T$ curves measured in zero magnetic field for all five single crystals.}
\end{figure}

To study the effect of temperature on the relationship between \Rff~and \jc~ given by Eq. (\ref{Eq1}), we performed $E$-$j$ measurements on all five samples of \BaCo~studied here over the temperature range $0.85~T_{c0}\leq T\leq 0.98~T_{c0}$ and extracted \Rff$(T)$ and $j_c(T)$. We show plots of $\rho_{\rm ff}^{-1}$ vs $j_c$ data measured at different $T$ and $x$ values in Figs.~\ref{f5}. The linear relationship between $\rho_{\rm ff}^{-1}$ and $j_{c}$ holds for all the dopings [with a positive slope $1/E_0$, see Eq. (\ref{Eq1})], implying that all these samples have negligible magnetoresistivity close to $T_c$. Moreover, the slopes $1/E_0$ of these plots are independent of temperature for the under-doped ($x=0.044$ and 0.056) and over-doped ($x=0.100$) samples since all the $\rho_{\rm ff}^{-1}$ vs $j_{c}$ data for different temperatures overlap. However, the slope $1/E_0$ increases slightly (strongly) with increasing temperature for the $x=0.060$ ($x=0.072$) single crystals. 
Notice that these two Co-doped samples are in the optimum-doping region of the phase diagram (see Fig.~\ref{f6}). It is reasonable to conclude that a phase or phase boundary that is sensitive to changes in temperature exists around the optimal doping and it affects the flux-flow behavior. Unfortunately, our data do not allow us to identify the precise nature of this additional phase. However, recent theoretical and experimental studies have shown that around optimal Co concentration there is a crossover between the SDW and SC coexisting phases, and the pure SC phase \cite{FernandesPRB2010, Huang2016}. Hence, the additional phase with pronounced temperature dependence revealed by our data may well be the SDW phase, although either spatially inhomogeneous microemulsion \cite{ChowdhuryPRB2015} or spin-glass \cite{CurroPRL2013} phases cannot be ruled out. 

\section{Conclusion}
We carried out current-voltage measurements as a function of temperature and applied magnetic field on \BaCo~single crystals with five different Co concentrations in the range $0.044 \leq x \leq 0.100$. We compare and contrast these results with results obtained by plotting published data on the canonical single band conventional superconductor Nb and multiband unconventional superconductor CeCoIn$_5$. 

We find that \jc~decreases with increasing $H$ in all these three systems, indicating that the de-pinning mechanism is the same in these conventional and unconventional superconductors. On the other hand, \Rff~shows opposite field dependences for the conventional Nb and unconventional  \BaCo~ and CeCoIn$_5$:  \Rff~increases with increasing $H$ in Nb, while it decreases sharply with increasing $H$ in  \BaCo~ and CeCoIn$_5$. The former result is typical of conventional type-II superconductors. Since in the flux-flow regime the dominant contribution to \Rff~ comes from the quasiparticle scattering in/around the vortex cores, the sharp increase in \Rff~ at low fields indicates the presence of a strong scattering mechanism in the unconventional superconductors \BaCo~ and CeCoIn$_5$, as discussed previously \cite{TaoPRL, Huang2016}.

We also revealed a relationship between the critical current density and flux-flow resistivity in all three systems studied: $\rho_{\rm ff}^{-1} = (j_0/E_0)\cdot(1- j_{c}/j_{0})$, where $E_0/j_0$ is the normal-state resistivity just above $T_c$ and $1/E_0$ is the slope of \Rff$^{-1}$ vs $j_c$. The above relationship is a consequence of the negligible magnetoresistance just above $T_c$ in all these three systems. The parameter $E_0$ is positive for all five Co concentrations studied and for CeCoIn$_5$, reflecting the abnormal increases of \Rff~ with decreasing $H$ and, as expected, it is negative for Nb. In addition, $E_0$ is temperature independent in under-doped and over-doped single crystals of  \BaCo, while $E_0$ decreases with increasing temperature for the single crystals around the optimal Co concentration ($ 0.060\leq x\leq 0.072 $).  This latter result reflects the presence of a temperature-dependent secondary, to superconductivity, phase, most likely the spin density wave phase, although either spatially
inhomogeneous micro-emulsion \cite{ChowdhuryPRB2015} or spin-glass \cite{CurroPRL2013} phases cannot be ruled out. This work has also shown that $I$-$V$ is a powerful transport technique that can probe the presence of different phases in  the superconducting state.

\textbf{Acknowledgments} 
We gratefully acknowledge Alan Baldwin's assistance with electronics and software development. This work was supported at KSU by the National Science Foundation grant Nos. DMR-1505826 and DMR-1506547. T. H.  acknowledges the support of NSFC, grant No. 11574338, and XDB04040300. H. X. acknowledges the support of NSFC, grant No. U1530402.

\bibliography{jc}

\begin{thebibliography}{35}%
\makeatletter
\providecommand \@ifxundefined [1]{%
 \@ifx{#1\undefined}
}%
\providecommand \@ifnum [1]{%
 \ifnum #1\expandafter \@firstoftwo
 \else \expandafter \@secondoftwo
 \fi
}%
\providecommand \@ifx [1]{%
 \ifx #1\expandafter \@firstoftwo
 \else \expandafter \@secondoftwo
 \fi
}%
\providecommand \natexlab [1]{#1}%
\providecommand \enquote  [1]{``#1''}%
\providecommand \bibnamefont  [1]{#1}%
\providecommand \bibfnamefont [1]{#1}%
\providecommand \citenamefont [1]{#1}%
\providecommand \href@noop [0]{\@secondoftwo}%
\providecommand \href [0]{\begingroup \@sanitize@url \@href}%
\providecommand \@href[1]{\@@startlink{#1}\@@href}%
\providecommand \@@href[1]{\endgroup#1\@@endlink}%
\providecommand \@sanitize@url [0]{\catcode `\\12\catcode `\$12\catcode
  `\&12\catcode `\#12\catcode `\^12\catcode `\_12\catcode `\%12\relax}%
\providecommand \@@startlink[1]{}%
\providecommand \@@endlink[0]{}%
\providecommand \url  [0]{\begingroup\@sanitize@url \@url }%
\providecommand \@url [1]{\endgroup\@href {#1}{\urlprefix }}%
\providecommand \urlprefix  [0]{URL }%
\providecommand \Eprint [0]{\href }%
\providecommand \doibase [0]{http://dx.doi.org/}%
\providecommand \selectlanguage [0]{\@gobble}%
\providecommand \bibinfo  [0]{\@secondoftwo}%
\providecommand \bibfield  [0]{\@secondoftwo}%
\providecommand \translation [1]{[#1]}%
\providecommand \BibitemOpen [0]{}%
\providecommand \bibitemStop [0]{}%
\providecommand \bibitemNoStop [0]{.\EOS\space}%
\providecommand \EOS [0]{\spacefactor3000\relax}%
\providecommand \BibitemShut  [1]{\csname bibitem#1\endcsname}%
\let\auto@bib@innerbib\@empty
\bibitem [{\citenamefont {Fernandes}\ \emph {et~al.}(2010)\citenamefont
  {Fernandes}, \citenamefont {Pratt}, \citenamefont {Tian}, \citenamefont
  {Zarestky}, \citenamefont {Kreyssig}, \citenamefont {Nandi}, \citenamefont
  {Kim}, \citenamefont {Thaler}, \citenamefont {Ni}, \citenamefont {Canfield},
  \citenamefont {McQueeney}, \citenamefont {Schmalian},\ and\ \citenamefont
  {Goldman}}]{FernandesPRB2010}%
  \BibitemOpen
  \bibfield  {author} {\bibinfo {author} {\bibfnamefont {R.~M.}\ \bibnamefont
  {Fernandes}}, \bibinfo {author} {\bibfnamefont {D.~K.}\ \bibnamefont
  {Pratt}}, \bibinfo {author} {\bibfnamefont {W.}~\bibnamefont {Tian}},
  \bibinfo {author} {\bibfnamefont {J.}~\bibnamefont {Zarestky}}, \bibinfo
  {author} {\bibfnamefont {A.}~\bibnamefont {Kreyssig}}, \bibinfo {author}
  {\bibfnamefont {S.}~\bibnamefont {Nandi}}, \bibinfo {author} {\bibfnamefont
  {M.~G.}\ \bibnamefont {Kim}}, \bibinfo {author} {\bibfnamefont
  {A.}~\bibnamefont {Thaler}}, \bibinfo {author} {\bibfnamefont
  {N.}~\bibnamefont {Ni}}, \bibinfo {author} {\bibfnamefont {P.~C.}\
  \bibnamefont {Canfield}}, \bibinfo {author} {\bibfnamefont {R.~J.}\
  \bibnamefont {McQueeney}}, \bibinfo {author} {\bibfnamefont {J.}~\bibnamefont
  {Schmalian}}, \ and\ \bibinfo {author} {\bibfnamefont {A.~I.}\ \bibnamefont
  {Goldman}},\ }\href {\doibase 10.1103/PhysRevB.81.140501} {\bibfield
  {journal} {\bibinfo  {journal} {Phys. Rev. B}\ }\textbf {\bibinfo {volume}
  {81}},\ \bibinfo {pages} {140501} (\bibinfo {year} {2010})}\BibitemShut
  {NoStop}%
\bibitem [{\citenamefont {Rotundu}\ and\ \citenamefont
  {Birgeneau}(2011)}]{RotunduPRB2011Secondorder}%
  \BibitemOpen
  \bibfield  {author} {\bibinfo {author} {\bibfnamefont {C.~R.}\ \bibnamefont
  {Rotundu}}\ and\ \bibinfo {author} {\bibfnamefont {R.~J.}\ \bibnamefont
  {Birgeneau}},\ }\href {\doibase 10.1103/PhysRevB.84.092501} {\bibfield
  {journal} {\bibinfo  {journal} {Phys. Rev. B}\ }\textbf {\bibinfo {volume}
  {84}},\ \bibinfo {pages} {092501} (\bibinfo {year} {2011})}\BibitemShut
  {NoStop}%
\bibitem [{\citenamefont {Chu}\ \emph {et~al.}(2010)\citenamefont {Chu},
  \citenamefont {Analytis}, \citenamefont {De~Greve}, \citenamefont {McMahon},
  \citenamefont {Islam}, \citenamefont {Yamamoto},\ and\ \citenamefont
  {Fisher}}]{Chu13082010}%
  \BibitemOpen
  \bibfield  {author} {\bibinfo {author} {\bibfnamefont {J.-H.}\ \bibnamefont
  {Chu}}, \bibinfo {author} {\bibfnamefont {J.~G.}\ \bibnamefont {Analytis}},
  \bibinfo {author} {\bibfnamefont {K.}~\bibnamefont {De~Greve}}, \bibinfo
  {author} {\bibfnamefont {P.~L.}\ \bibnamefont {McMahon}}, \bibinfo {author}
  {\bibfnamefont {Z.}~\bibnamefont {Islam}}, \bibinfo {author} {\bibfnamefont
  {Y.}~\bibnamefont {Yamamoto}}, \ and\ \bibinfo {author} {\bibfnamefont
  {I.~R.}\ \bibnamefont {Fisher}},\ }\href {\doibase 10.1126/science.1190482}
  {\bibfield  {journal} {\bibinfo  {journal} {Science}\ }\textbf {\bibinfo
  {volume} {329}},\ \bibinfo {pages} {824} (\bibinfo {year}
  {2010})}\BibitemShut {NoStop}%
\bibitem [{\citenamefont {Chu}\ \emph {et~al.}(2012)\citenamefont {Chu},
  \citenamefont {Kuo}, \citenamefont {Analytis},\ and\ \citenamefont
  {Fisher}}]{Chu10082012}%
  \BibitemOpen
  \bibfield  {author} {\bibinfo {author} {\bibfnamefont {J.-H.}\ \bibnamefont
  {Chu}}, \bibinfo {author} {\bibfnamefont {H.-H.}\ \bibnamefont {Kuo}},
  \bibinfo {author} {\bibfnamefont {J.~G.}\ \bibnamefont {Analytis}}, \ and\
  \bibinfo {author} {\bibfnamefont {I.~R.}\ \bibnamefont {Fisher}},\ }\href
  {\doibase 10.1126/science.1221713} {\bibfield  {journal} {\bibinfo  {journal}
  {Science}\ }\textbf {\bibinfo {volume} {337}},\ \bibinfo {pages} {710}
  (\bibinfo {year} {2012})}\BibitemShut {NoStop}%
\bibitem [{\citenamefont {Dioguardi}\ \emph {et~al.}(2013)\citenamefont
  {Dioguardi}, \citenamefont {Crocker}, \citenamefont {Shockley}, \citenamefont
  {Lin}, \citenamefont {Shirer}, \citenamefont {Nisson}, \citenamefont
  {Lawson}, \citenamefont {apRoberts Warren}, \citenamefont {Canfield},
  \citenamefont {Bud'ko}, \citenamefont {Ran},\ and\ \citenamefont
  {Curro}}]{CurroPRL2013}%
  \BibitemOpen
  \bibfield  {author} {\bibinfo {author} {\bibfnamefont {A.~P.}\ \bibnamefont
  {Dioguardi}}, \bibinfo {author} {\bibfnamefont {J.}~\bibnamefont {Crocker}},
  \bibinfo {author} {\bibfnamefont {A.~C.}\ \bibnamefont {Shockley}}, \bibinfo
  {author} {\bibfnamefont {C.~H.}\ \bibnamefont {Lin}}, \bibinfo {author}
  {\bibfnamefont {K.~R.}\ \bibnamefont {Shirer}}, \bibinfo {author}
  {\bibfnamefont {D.~M.}\ \bibnamefont {Nisson}}, \bibinfo {author}
  {\bibfnamefont {M.~M.}\ \bibnamefont {Lawson}}, \bibinfo {author}
  {\bibfnamefont {N.}~\bibnamefont {apRoberts Warren}}, \bibinfo {author}
  {\bibfnamefont {P.~C.}\ \bibnamefont {Canfield}}, \bibinfo {author}
  {\bibfnamefont {S.~L.}\ \bibnamefont {Bud'ko}}, \bibinfo {author}
  {\bibfnamefont {S.}~\bibnamefont {Ran}}, \ and\ \bibinfo {author}
  {\bibfnamefont {N.~J.}\ \bibnamefont {Curro}},\ }\href {\doibase
  10.1103/PhysRevLett.111.207201} {\bibfield  {journal} {\bibinfo  {journal}
  {Phys. Rev. Lett.}\ }\textbf {\bibinfo {volume} {111}},\ \bibinfo {pages}
  {207201} (\bibinfo {year} {2013})}\BibitemShut {NoStop}%
\bibitem [{\citenamefont {Ning}\ \emph {et~al.}(2009)\citenamefont {Ning},
  \citenamefont {Ahilan}, \citenamefont {Imai}, \citenamefont {Sefat},
  \citenamefont {Jin}, \citenamefont {McGuire}, \citenamefont {Sales},\ and\
  \citenamefont {Mandrus}}]{FanlongNingJPSJ2009}%
  \BibitemOpen
  \bibfield  {author} {\bibinfo {author} {\bibfnamefont {F.}~\bibnamefont
  {Ning}}, \bibinfo {author} {\bibfnamefont {K.}~\bibnamefont {Ahilan}},
  \bibinfo {author} {\bibfnamefont {T.}~\bibnamefont {Imai}}, \bibinfo {author}
  {\bibfnamefont {A.~S.}\ \bibnamefont {Sefat}}, \bibinfo {author}
  {\bibfnamefont {R.}~\bibnamefont {Jin}}, \bibinfo {author} {\bibfnamefont
  {M.~A.}\ \bibnamefont {McGuire}}, \bibinfo {author} {\bibfnamefont {B.~C.}\
  \bibnamefont {Sales}}, \ and\ \bibinfo {author} {\bibfnamefont
  {D.}~\bibnamefont {Mandrus}},\ }\href {\doibase 10.1143/JPSJ.78.013711}
  {\bibfield  {journal} {\bibinfo  {journal} {Journal of the Physical Society
  of Japan}\ }\textbf {\bibinfo {volume} {78}},\ \bibinfo {pages} {013711}
  (\bibinfo {year} {2009})}\BibitemShut {NoStop}%
\bibitem [{\citenamefont {Fernandes}\ \emph {et~al.}(2013)\citenamefont
  {Fernandes}, \citenamefont {Maiti}, \citenamefont {W\"olfle},\ and\
  \citenamefont {Chubukov}}]{fernandesPRL2013howmanyQCP}%
  \BibitemOpen
  \bibfield  {author} {\bibinfo {author} {\bibfnamefont {R.~M.}\ \bibnamefont
  {Fernandes}}, \bibinfo {author} {\bibfnamefont {S.}~\bibnamefont {Maiti}},
  \bibinfo {author} {\bibfnamefont {P.}~\bibnamefont {W\"olfle}}, \ and\
  \bibinfo {author} {\bibfnamefont {A.~V.}\ \bibnamefont {Chubukov}},\ }\href
  {\doibase 10.1103/PhysRevLett.111.057001} {\bibfield  {journal} {\bibinfo
  {journal} {Phys. Rev. Lett.}\ }\textbf {\bibinfo {volume} {111}},\ \bibinfo
  {pages} {057001} (\bibinfo {year} {2013})}\BibitemShut {NoStop}%
\bibitem [{\citenamefont {Arsenijevi\ifmmode~\acute{c}\else \'{c}\fi{}}\ \emph
  {et~al.}(2013)\citenamefont {Arsenijevi\ifmmode~\acute{c}\else \'{c}\fi{}},
  \citenamefont {Hodovanets}, \citenamefont {Ga\'al}, \citenamefont {Forr\'o},
  \citenamefont {Bud'ko},\ and\ \citenamefont {Canfield}}]{ThermPowerQCP}%
  \BibitemOpen
  \bibfield  {author} {\bibinfo {author} {\bibfnamefont {S.}~\bibnamefont
  {Arsenijevi\ifmmode~\acute{c}\else \'{c}\fi{}}}, \bibinfo {author}
  {\bibfnamefont {H.}~\bibnamefont {Hodovanets}}, \bibinfo {author}
  {\bibfnamefont {R.}~\bibnamefont {Ga\'al}}, \bibinfo {author} {\bibfnamefont
  {L.}~\bibnamefont {Forr\'o}}, \bibinfo {author} {\bibfnamefont {S.~L.}\
  \bibnamefont {Bud'ko}}, \ and\ \bibinfo {author} {\bibfnamefont {P.~C.}\
  \bibnamefont {Canfield}},\ }\href {\doibase 10.1103/PhysRevB.87.224508}
  {\bibfield  {journal} {\bibinfo  {journal} {Phys. Rev. B}\ }\textbf {\bibinfo
  {volume} {87}},\ \bibinfo {pages} {224508} (\bibinfo {year}
  {2013})}\BibitemShut {NoStop}%
\bibitem [{\citenamefont {Hu}\ \emph {et~al.}(2015)\citenamefont {Hu},
  \citenamefont {Lu}, \citenamefont {Zhang}, \citenamefont {Luo}, \citenamefont
  {Li}, \citenamefont {Wang}, \citenamefont {Chen}, \citenamefont {Han},
  \citenamefont {Banjara}, \citenamefont {Sapkota}, \citenamefont {Kreyssig},
  \citenamefont {Goldman}, \citenamefont {Yamani}, \citenamefont {Niedermayer},
  \citenamefont {Skoulatos}, \citenamefont {Georgii}, \citenamefont {Keller},
  \citenamefont {Wang}, \citenamefont {Yu},\ and\ \citenamefont
  {Dai}}]{HuDingPRL2015}%
  \BibitemOpen
  \bibfield  {author} {\bibinfo {author} {\bibfnamefont {D.}~\bibnamefont
  {Hu}}, \bibinfo {author} {\bibfnamefont {X.}~\bibnamefont {Lu}}, \bibinfo
  {author} {\bibfnamefont {W.}~\bibnamefont {Zhang}}, \bibinfo {author}
  {\bibfnamefont {H.}~\bibnamefont {Luo}}, \bibinfo {author} {\bibfnamefont
  {S.}~\bibnamefont {Li}}, \bibinfo {author} {\bibfnamefont {P.}~\bibnamefont
  {Wang}}, \bibinfo {author} {\bibfnamefont {G.}~\bibnamefont {Chen}}, \bibinfo
  {author} {\bibfnamefont {F.}~\bibnamefont {Han}}, \bibinfo {author}
  {\bibfnamefont {S.~R.}\ \bibnamefont {Banjara}}, \bibinfo {author}
  {\bibfnamefont {A.}~\bibnamefont {Sapkota}}, \bibinfo {author} {\bibfnamefont
  {A.}~\bibnamefont {Kreyssig}}, \bibinfo {author} {\bibfnamefont {A.~I.}\
  \bibnamefont {Goldman}}, \bibinfo {author} {\bibfnamefont {Z.}~\bibnamefont
  {Yamani}}, \bibinfo {author} {\bibfnamefont {C.}~\bibnamefont {Niedermayer}},
  \bibinfo {author} {\bibfnamefont {M.}~\bibnamefont {Skoulatos}}, \bibinfo
  {author} {\bibfnamefont {R.}~\bibnamefont {Georgii}}, \bibinfo {author}
  {\bibfnamefont {T.}~\bibnamefont {Keller}}, \bibinfo {author} {\bibfnamefont
  {P.}~\bibnamefont {Wang}}, \bibinfo {author} {\bibfnamefont {W.}~\bibnamefont
  {Yu}}, \ and\ \bibinfo {author} {\bibfnamefont {P.}~\bibnamefont {Dai}},\
  }\href {\doibase 10.1103/PhysRevLett.114.157002} {\bibfield  {journal}
  {\bibinfo  {journal} {Phys. Rev. Lett.}\ }\textbf {\bibinfo {volume} {114}},\
  \bibinfo {pages} {157002} (\bibinfo {year} {2015})}\BibitemShut {NoStop}%
\bibitem [{\citenamefont {Chowdhury}\ \emph {et~al.}(2015)\citenamefont
  {Chowdhury}, \citenamefont {Orenstein}, \citenamefont {Sachdev},\ and\
  \citenamefont {Senthil}}]{ChowdhuryPRB2015}%
  \BibitemOpen
  \bibfield  {author} {\bibinfo {author} {\bibfnamefont {D.}~\bibnamefont
  {Chowdhury}}, \bibinfo {author} {\bibfnamefont {J.}~\bibnamefont
  {Orenstein}}, \bibinfo {author} {\bibfnamefont {S.}~\bibnamefont {Sachdev}},
  \ and\ \bibinfo {author} {\bibfnamefont {T.}~\bibnamefont {Senthil}},\ }\href
  {\doibase 10.1103/PhysRevB.92.081113} {\bibfield  {journal} {\bibinfo
  {journal} {Phys. Rev. B}\ }\textbf {\bibinfo {volume} {92}},\ \bibinfo
  {pages} {081113} (\bibinfo {year} {2015})}\BibitemShut {NoStop}%
\bibitem [{\citenamefont {Dew-Hughes}(1974)}]{D.Hughes1974}%
  \BibitemOpen
  \bibfield  {author} {\bibinfo {author} {\bibfnamefont {D.}~\bibnamefont
  {Dew-Hughes}},\ }\href {\doibase 10.1080/14786439808206556} {\bibfield
  {journal} {\bibinfo  {journal} {Philosophical Magazine}\ }\textbf {\bibinfo
  {volume} {30}},\ \bibinfo {pages} {293} (\bibinfo {year} {1974})}\BibitemShut
  {NoStop}%
\bibitem [{\citenamefont {Strnad}\ \emph {et~al.}(1964)\citenamefont {Strnad},
  \citenamefont {Hempstead},\ and\ \citenamefont {Kim}}]{Y.B.Kim1964}%
  \BibitemOpen
  \bibfield  {author} {\bibinfo {author} {\bibfnamefont {A.~R.}\ \bibnamefont
  {Strnad}}, \bibinfo {author} {\bibfnamefont {C.~F.}\ \bibnamefont
  {Hempstead}}, \ and\ \bibinfo {author} {\bibfnamefont {Y.~B.}\ \bibnamefont
  {Kim}},\ }\href {\doibase 10.1103/PhysRevLett.13.794} {\bibfield  {journal}
  {\bibinfo  {journal} {Phys. Rev. Lett.}\ }\textbf {\bibinfo {volume} {13}},\
  \bibinfo {pages} {794} (\bibinfo {year} {1964})}\BibitemShut {NoStop}%
\bibitem [{\citenamefont {Kim}\ \emph {et~al.}(1965)\citenamefont {Kim},
  \citenamefont {Hempstead},\ and\ \citenamefont {Strnad}}]{Y.B.Kim1965}%
  \BibitemOpen
  \bibfield  {author} {\bibinfo {author} {\bibfnamefont {Y.~B.}\ \bibnamefont
  {Kim}}, \bibinfo {author} {\bibfnamefont {C.~F.}\ \bibnamefont {Hempstead}},
  \ and\ \bibinfo {author} {\bibfnamefont {A.~R.}\ \bibnamefont {Strnad}},\
  }\href {\doibase 10.1103/PhysRev.139.A1163} {\bibfield  {journal} {\bibinfo
  {journal} {Phys. Rev.}\ }\textbf {\bibinfo {volume} {139}},\ \bibinfo {pages}
  {A1163} (\bibinfo {year} {1965})}\BibitemShut {NoStop}%
\bibitem [{\citenamefont {Prozorov}\ \emph {et~al.}(2008)\citenamefont
  {Prozorov}, \citenamefont {Ni}, \citenamefont {Tanatar}, \citenamefont
  {Kogan}, \citenamefont {Gordon}, \citenamefont {Martin}, \citenamefont
  {Blomberg}, \citenamefont {Prommapan}, \citenamefont {Yan}, \citenamefont
  {Bud'ko},\ and\ \citenamefont {Canfield}}]{ProzorovPRB2008}%
  \BibitemOpen
  \bibfield  {author} {\bibinfo {author} {\bibfnamefont {R.}~\bibnamefont
  {Prozorov}}, \bibinfo {author} {\bibfnamefont {N.}~\bibnamefont {Ni}},
  \bibinfo {author} {\bibfnamefont {M.~A.}\ \bibnamefont {Tanatar}}, \bibinfo
  {author} {\bibfnamefont {V.~G.}\ \bibnamefont {Kogan}}, \bibinfo {author}
  {\bibfnamefont {R.~T.}\ \bibnamefont {Gordon}}, \bibinfo {author}
  {\bibfnamefont {C.}~\bibnamefont {Martin}}, \bibinfo {author} {\bibfnamefont
  {E.~C.}\ \bibnamefont {Blomberg}}, \bibinfo {author} {\bibfnamefont
  {P.}~\bibnamefont {Prommapan}}, \bibinfo {author} {\bibfnamefont {J.~Q.}\
  \bibnamefont {Yan}}, \bibinfo {author} {\bibfnamefont {S.~L.}\ \bibnamefont
  {Bud'ko}}, \ and\ \bibinfo {author} {\bibfnamefont {P.~C.}\ \bibnamefont
  {Canfield}},\ }\href {\doibase 10.1103/PhysRevB.78.224506} {\bibfield
  {journal} {\bibinfo  {journal} {Phys. Rev. B}\ }\textbf {\bibinfo {volume}
  {78}},\ \bibinfo {pages} {224506} (\bibinfo {year} {2008})}\BibitemShut
  {NoStop}%
\bibitem [{\citenamefont {Tanatar}\ \emph {et~al.}(2010)\citenamefont
  {Tanatar}, \citenamefont {Ni}, \citenamefont {Bud’ko}, \citenamefont
  {Canfield},\ and\ \citenamefont {Prozorov}}]{TanatarIOP2010}%
  \BibitemOpen
  \bibfield  {author} {\bibinfo {author} {\bibfnamefont {M.~A.}\ \bibnamefont
  {Tanatar}}, \bibinfo {author} {\bibfnamefont {N.}~\bibnamefont {Ni}},
  \bibinfo {author} {\bibfnamefont {S.~L.}\ \bibnamefont {Bud’ko}}, \bibinfo
  {author} {\bibfnamefont {P.~C.}\ \bibnamefont {Canfield}}, \ and\ \bibinfo
  {author} {\bibfnamefont {R.}~\bibnamefont {Prozorov}},\ }\href
  {http://stacks.iop.org/0953-2048/23/i=5/a=054002} {\bibfield  {journal}
  {\bibinfo  {journal} {Superconductor Science and Technology}\ }\textbf
  {\bibinfo {volume} {23}},\ \bibinfo {pages} {054002} (\bibinfo {year}
  {2010})}\BibitemShut {NoStop}%
\bibitem [{\citenamefont {Hanisch}\ \emph {et~al.}(2011)\citenamefont
  {Hanisch}, \citenamefont {Iida}, \citenamefont {Haindl}, \citenamefont
  {Kurth}, \citenamefont {Kauffmann}, \citenamefont {Kidszun}, \citenamefont
  {Thersleff}, \citenamefont {Freudenberger}, \citenamefont {Schultz},\ and\
  \citenamefont {Holzapfel}}]{J.HanischIEEE2011}%
  \BibitemOpen
  \bibfield  {author} {\bibinfo {author} {\bibfnamefont {J.}~\bibnamefont
  {Hanisch}}, \bibinfo {author} {\bibfnamefont {K.}~\bibnamefont {Iida}},
  \bibinfo {author} {\bibfnamefont {S.}~\bibnamefont {Haindl}}, \bibinfo
  {author} {\bibfnamefont {F.}~\bibnamefont {Kurth}}, \bibinfo {author}
  {\bibfnamefont {A.}~\bibnamefont {Kauffmann}}, \bibinfo {author}
  {\bibfnamefont {M.}~\bibnamefont {Kidszun}}, \bibinfo {author} {\bibfnamefont
  {T.}~\bibnamefont {Thersleff}}, \bibinfo {author} {\bibfnamefont
  {J.}~\bibnamefont {Freudenberger}}, \bibinfo {author} {\bibfnamefont
  {L.}~\bibnamefont {Schultz}}, \ and\ \bibinfo {author} {\bibfnamefont
  {B.}~\bibnamefont {Holzapfel}},\ }\href {\doibase 10.1109/TASC.2010.2100348}
  {\bibfield  {journal} {\bibinfo  {journal} {IEEE Transactions on Applied
  Superconductivity}\ }\textbf {\bibinfo {volume} {21}},\ \bibinfo {pages}
  {2887} (\bibinfo {year} {2011})}\BibitemShut {NoStop}%
\bibitem [{\citenamefont {Okada}\ \emph {et~al.}(2012)\citenamefont {Okada},
  \citenamefont {Takahashi}, \citenamefont {Imai}, \citenamefont {Kitagawa},
  \citenamefont {Matsubayashi}, \citenamefont {Uwatoko},\ and\ \citenamefont
  {Maeda}}]{MaedaPRB2012}%
  \BibitemOpen
  \bibfield  {author} {\bibinfo {author} {\bibfnamefont {T.}~\bibnamefont
  {Okada}}, \bibinfo {author} {\bibfnamefont {H.}~\bibnamefont {Takahashi}},
  \bibinfo {author} {\bibfnamefont {Y.}~\bibnamefont {Imai}}, \bibinfo {author}
  {\bibfnamefont {K.}~\bibnamefont {Kitagawa}}, \bibinfo {author}
  {\bibfnamefont {K.}~\bibnamefont {Matsubayashi}}, \bibinfo {author}
  {\bibfnamefont {Y.}~\bibnamefont {Uwatoko}}, \ and\ \bibinfo {author}
  {\bibfnamefont {A.}~\bibnamefont {Maeda}},\ }\href {\doibase
  10.1103/PhysRevB.86.064516} {\bibfield  {journal} {\bibinfo  {journal} {Phys.
  Rev. B}\ }\textbf {\bibinfo {volume} {86}},\ \bibinfo {pages} {064516}
  (\bibinfo {year} {2012})}\BibitemShut {NoStop}%
\bibitem [{\citenamefont {Okada}\ \emph {et~al.}(2013)\citenamefont {Okada},
  \citenamefont {Takahashi}, \citenamefont {Imai}, \citenamefont {Kitagawa},
  \citenamefont {Matsubayashi}, \citenamefont {Uwatoko},\ and\ \citenamefont
  {Maeda}}]{MaedaPhyC2013}%
  \BibitemOpen
  \bibfield  {author} {\bibinfo {author} {\bibfnamefont {T.}~\bibnamefont
  {Okada}}, \bibinfo {author} {\bibfnamefont {H.}~\bibnamefont {Takahashi}},
  \bibinfo {author} {\bibfnamefont {Y.}~\bibnamefont {Imai}}, \bibinfo {author}
  {\bibfnamefont {K.}~\bibnamefont {Kitagawa}}, \bibinfo {author}
  {\bibfnamefont {K.}~\bibnamefont {Matsubayashi}}, \bibinfo {author}
  {\bibfnamefont {Y.}~\bibnamefont {Uwatoko}}, \ and\ \bibinfo {author}
  {\bibfnamefont {A.}~\bibnamefont {Maeda}},\ }\href {\doibase
  http://dx.doi.org/10.1016/j.physc.2013.04.074} {\bibfield  {journal}
  {\bibinfo  {journal} {Physica C: Superconductivity}\ }\textbf {\bibinfo
  {volume} {494}},\ \bibinfo {pages} {109 } (\bibinfo {year} {2013})},\
  \bibinfo {note} {proceedings of the 25th International Symposium on
  Superconductivity (ISS 2012) Advances in Superconductivity
  \{XXV\}}\BibitemShut {NoStop}%
\bibitem [{\citenamefont {Okada}\ \emph {et~al.}(2014)\citenamefont {Okada},
  \citenamefont {Imai}, \citenamefont {Takahashi}, \citenamefont {Nakajima},
  \citenamefont {Iyo}, \citenamefont {Eisaki},\ and\ \citenamefont
  {Maeda}}]{MaedaPhyC2014}%
  \BibitemOpen
  \bibfield  {author} {\bibinfo {author} {\bibfnamefont {T.}~\bibnamefont
  {Okada}}, \bibinfo {author} {\bibfnamefont {Y.}~\bibnamefont {Imai}},
  \bibinfo {author} {\bibfnamefont {H.}~\bibnamefont {Takahashi}}, \bibinfo
  {author} {\bibfnamefont {M.}~\bibnamefont {Nakajima}}, \bibinfo {author}
  {\bibfnamefont {A.}~\bibnamefont {Iyo}}, \bibinfo {author} {\bibfnamefont
  {H.}~\bibnamefont {Eisaki}}, \ and\ \bibinfo {author} {\bibfnamefont
  {A.}~\bibnamefont {Maeda}},\ }\href {\doibase
  http://dx.doi.org/10.1016/j.physc.2014.03.025} {\bibfield  {journal}
  {\bibinfo  {journal} {Physica C: Superconductivity}\ }\textbf {\bibinfo
  {volume} {504}},\ \bibinfo {pages} {24 } (\bibinfo {year} {2014})},\ \bibinfo
  {note} {proceedings of the 26th International Symposium on
  Superconductivity}\BibitemShut {NoStop}%
\bibitem [{\citenamefont {Okada}\ \emph {et~al.}(2015)\citenamefont {Okada},
  \citenamefont {Nabeshima}, \citenamefont {Takahashi}, \citenamefont {Imai},\
  and\ \citenamefont {Maeda}}]{MaedaPRB2015}%
  \BibitemOpen
  \bibfield  {author} {\bibinfo {author} {\bibfnamefont {T.}~\bibnamefont
  {Okada}}, \bibinfo {author} {\bibfnamefont {F.}~\bibnamefont {Nabeshima}},
  \bibinfo {author} {\bibfnamefont {H.}~\bibnamefont {Takahashi}}, \bibinfo
  {author} {\bibfnamefont {Y.}~\bibnamefont {Imai}}, \ and\ \bibinfo {author}
  {\bibfnamefont {A.}~\bibnamefont {Maeda}},\ }\href {\doibase
  10.1103/PhysRevB.91.054510} {\bibfield  {journal} {\bibinfo  {journal} {Phys.
  Rev. B}\ }\textbf {\bibinfo {volume} {91}},\ \bibinfo {pages} {054510}
  (\bibinfo {year} {2015})}\BibitemShut {NoStop}%
\bibitem [{\citenamefont {Chu}\ \emph {et~al.}(2009)\citenamefont {Chu},
  \citenamefont {Analytis}, \citenamefont {Kucharczyk},\ and\ \citenamefont
  {Fisher}}]{ChuPRB2009}%
  \BibitemOpen
  \bibfield  {author} {\bibinfo {author} {\bibfnamefont {J.-H.}\ \bibnamefont
  {Chu}}, \bibinfo {author} {\bibfnamefont {J.~G.}\ \bibnamefont {Analytis}},
  \bibinfo {author} {\bibfnamefont {C.}~\bibnamefont {Kucharczyk}}, \ and\
  \bibinfo {author} {\bibfnamefont {I.~R.}\ \bibnamefont {Fisher}},\ }\href
  {\doibase 10.1103/PhysRevB.79.014506} {\bibfield  {journal} {\bibinfo
  {journal} {Phys. Rev. B}\ }\textbf {\bibinfo {volume} {79}},\ \bibinfo
  {pages} {014506} (\bibinfo {year} {2009})}\BibitemShut {NoStop}%
\bibitem [{\citenamefont {Ni}\ \emph {et~al.}(2008)\citenamefont {Ni},
  \citenamefont {Tillman}, \citenamefont {Yan}, \citenamefont {Kracher},
  \citenamefont {Hannahs}, \citenamefont {Bud'ko},\ and\ \citenamefont
  {Canfield}}]{NiNiPRB2008EffectofCo}%
  \BibitemOpen
  \bibfield  {author} {\bibinfo {author} {\bibfnamefont {N.}~\bibnamefont
  {Ni}}, \bibinfo {author} {\bibfnamefont {M.~E.}\ \bibnamefont {Tillman}},
  \bibinfo {author} {\bibfnamefont {J.-Q.}\ \bibnamefont {Yan}}, \bibinfo
  {author} {\bibfnamefont {A.}~\bibnamefont {Kracher}}, \bibinfo {author}
  {\bibfnamefont {S.~T.}\ \bibnamefont {Hannahs}}, \bibinfo {author}
  {\bibfnamefont {S.~L.}\ \bibnamefont {Bud'ko}}, \ and\ \bibinfo {author}
  {\bibfnamefont {P.~C.}\ \bibnamefont {Canfield}},\ }\href {\doibase
  10.1103/PhysRevB.78.214515} {\bibfield  {journal} {\bibinfo  {journal} {Phys.
  Rev. B}\ }\textbf {\bibinfo {volume} {78}},\ \bibinfo {pages} {214515}
  (\bibinfo {year} {2008})}\BibitemShut {NoStop}%
\bibitem [{\citenamefont {Reid}\ \emph {et~al.}(2010)\citenamefont {Reid},
  \citenamefont {Tanatar}, \citenamefont {Luo}, \citenamefont {Shakeripour},
  \citenamefont {Doiron-Leyraud}, \citenamefont {Ni}, \citenamefont {Bud'ko},
  \citenamefont {Canfield}, \citenamefont {Prozorov},\ and\ \citenamefont
  {Taillefer}}]{J.Reid_PRB2010}%
  \BibitemOpen
  \bibfield  {author} {\bibinfo {author} {\bibfnamefont {J.-P.}\ \bibnamefont
  {Reid}}, \bibinfo {author} {\bibfnamefont {M.~A.}\ \bibnamefont {Tanatar}},
  \bibinfo {author} {\bibfnamefont {X.~G.}\ \bibnamefont {Luo}}, \bibinfo
  {author} {\bibfnamefont {H.}~\bibnamefont {Shakeripour}}, \bibinfo {author}
  {\bibfnamefont {N.}~\bibnamefont {Doiron-Leyraud}}, \bibinfo {author}
  {\bibfnamefont {N.}~\bibnamefont {Ni}}, \bibinfo {author} {\bibfnamefont
  {S.~L.}\ \bibnamefont {Bud'ko}}, \bibinfo {author} {\bibfnamefont {P.~C.}\
  \bibnamefont {Canfield}}, \bibinfo {author} {\bibfnamefont {R.}~\bibnamefont
  {Prozorov}}, \ and\ \bibinfo {author} {\bibfnamefont {L.}~\bibnamefont
  {Taillefer}},\ }\href {\doibase 10.1103/PhysRevB.82.064501} {\bibfield
  {journal} {\bibinfo  {journal} {Phys. Rev. B}\ }\textbf {\bibinfo {volume}
  {82}},\ \bibinfo {pages} {064501} (\bibinfo {year} {2010})}\BibitemShut
  {NoStop}%
\bibitem [{\citenamefont {Bardeen}\ and\ \citenamefont
  {Stephen}(1965)}]{BardeenStephen1965}%
  \BibitemOpen
  \bibfield  {author} {\bibinfo {author} {\bibfnamefont {J.}~\bibnamefont
  {Bardeen}}\ and\ \bibinfo {author} {\bibfnamefont {M.~J.}\ \bibnamefont
  {Stephen}},\ }\href {\doibase 10.1103/PhysRev.140.A1197} {\bibfield
  {journal} {\bibinfo  {journal} {Phys. Rev.}\ }\textbf {\bibinfo {volume}
  {140}},\ \bibinfo {pages} {A1197} (\bibinfo {year} {1965})}\BibitemShut
  {NoStop}%
\bibitem [{\citenamefont {Kopnin}\ and\ \citenamefont
  {Lopatin}(1995)}]{KopninPRB1995}%
  \BibitemOpen
  \bibfield  {author} {\bibinfo {author} {\bibfnamefont {N.~B.}\ \bibnamefont
  {Kopnin}}\ and\ \bibinfo {author} {\bibfnamefont {A.~V.}\ \bibnamefont
  {Lopatin}},\ }\href {\doibase 10.1103/PhysRevB.51.15291} {\bibfield
  {journal} {\bibinfo  {journal} {Phys. Rev. B}\ }\textbf {\bibinfo {volume}
  {51}},\ \bibinfo {pages} {15291} (\bibinfo {year} {1995})}\BibitemShut
  {NoStop}%
\bibitem [{\citenamefont {Kopnin}\ and\ \citenamefont
  {Volovik}(1997)}]{KoppinPRL1997}%
  \BibitemOpen
  \bibfield  {author} {\bibinfo {author} {\bibfnamefont {N.~B.}\ \bibnamefont
  {Kopnin}}\ and\ \bibinfo {author} {\bibfnamefont {G.~E.}\ \bibnamefont
  {Volovik}},\ }\href {\doibase 10.1103/PhysRevLett.79.1377} {\bibfield
  {journal} {\bibinfo  {journal} {Phys. Rev. Lett.}\ }\textbf {\bibinfo
  {volume} {79}},\ \bibinfo {pages} {1377} (\bibinfo {year}
  {1997})}\BibitemShut {NoStop}%
\bibitem [{\citenamefont {Kambe}\ \emph {et~al.}(1999)\citenamefont {Kambe},
  \citenamefont {Huxley}, \citenamefont {Rodi\`ere},\ and\ \citenamefont
  {Flouquet}}]{KambePRL1999UPt3}%
  \BibitemOpen
  \bibfield  {author} {\bibinfo {author} {\bibfnamefont {S.}~\bibnamefont
  {Kambe}}, \bibinfo {author} {\bibfnamefont {A.~D.}\ \bibnamefont {Huxley}},
  \bibinfo {author} {\bibfnamefont {P.}~\bibnamefont {Rodi\`ere}}, \ and\
  \bibinfo {author} {\bibfnamefont {J.}~\bibnamefont {Flouquet}},\ }\href
  {\doibase 10.1103/PhysRevLett.83.1842} {\bibfield  {journal} {\bibinfo
  {journal} {Phys. Rev. Lett.}\ }\textbf {\bibinfo {volume} {83}},\ \bibinfo
  {pages} {1842} (\bibinfo {year} {1999})}\BibitemShut {NoStop}%
\bibitem [{\citenamefont {Huebener}\ \emph {et~al.}(1970)\citenamefont
  {Huebener}, \citenamefont {Kampwirth},\ and\ \citenamefont
  {Seher}}]{Huebener1970}%
  \BibitemOpen
  \bibfield  {author} {\bibinfo {author} {\bibfnamefont {R.~P.}\ \bibnamefont
  {Huebener}}, \bibinfo {author} {\bibfnamefont {R.~T.}\ \bibnamefont
  {Kampwirth}}, \ and\ \bibinfo {author} {\bibfnamefont {A.}~\bibnamefont
  {Seher}},\ }\href {\doibase 10.1007/BF00628104} {\bibfield  {journal}
  {\bibinfo  {journal} {Journal of Low Temperature Physics}\ }\textbf {\bibinfo
  {volume} {2}},\ \bibinfo {pages} {113} (\bibinfo {year} {1970})}\BibitemShut
  {NoStop}%
\bibitem [{\citenamefont {Peroz}\ and\ \citenamefont
  {Villard}(2005)}]{C.PerozPRB2005}%
  \BibitemOpen
  \bibfield  {author} {\bibinfo {author} {\bibfnamefont {C.}~\bibnamefont
  {Peroz}}\ and\ \bibinfo {author} {\bibfnamefont {C.}~\bibnamefont
  {Villard}},\ }\href {\doibase 10.1103/PhysRevB.72.014515} {\bibfield
  {journal} {\bibinfo  {journal} {Phys. Rev. B}\ }\textbf {\bibinfo {volume}
  {72}},\ \bibinfo {pages} {014515} (\bibinfo {year} {2005})}\BibitemShut
  {NoStop}%
\bibitem [{\citenamefont {Hu}\ \emph {et~al.}(2012)\citenamefont {Hu},
  \citenamefont {Xiao}, \citenamefont {Sayles}, \citenamefont {Dzero},
  \citenamefont {Maple},\ and\ \citenamefont {Almasan}}]{TaoPRL}%
  \BibitemOpen
  \bibfield  {author} {\bibinfo {author} {\bibfnamefont {T.}~\bibnamefont
  {Hu}}, \bibinfo {author} {\bibfnamefont {H.}~\bibnamefont {Xiao}}, \bibinfo
  {author} {\bibfnamefont {T.~A.}\ \bibnamefont {Sayles}}, \bibinfo {author}
  {\bibfnamefont {M.}~\bibnamefont {Dzero}}, \bibinfo {author} {\bibfnamefont
  {M.~B.}\ \bibnamefont {Maple}}, \ and\ \bibinfo {author} {\bibfnamefont
  {C.~C.}\ \bibnamefont {Almasan}},\ }\href {\doibase
  10.1103/PhysRevLett.108.056401} {\bibfield  {journal} {\bibinfo  {journal}
  {Phys. Rev. Lett.}\ }\textbf {\bibinfo {volume} {108}},\ \bibinfo {pages}
  {056401} (\bibinfo {year} {2012})}\BibitemShut {NoStop}%
\bibitem [{\citenamefont {Huang}\ \emph {et~al.}(2017)\citenamefont {Huang},
  \citenamefont {Haney}, \citenamefont {Singh}, \citenamefont {Hu},
  \citenamefont {Xiao}, \citenamefont {Wen}, \citenamefont {Dzero},\ and\
  \citenamefont {Almasan}}]{Huang2016}%
  \BibitemOpen
  \bibfield  {author} {\bibinfo {author} {\bibfnamefont {X.~Y.}\ \bibnamefont
  {Huang}}, \bibinfo {author} {\bibfnamefont {D.~J.}\ \bibnamefont {Haney}},
  \bibinfo {author} {\bibfnamefont {Y.~P.}\ \bibnamefont {Singh}}, \bibinfo
  {author} {\bibfnamefont {T.}~\bibnamefont {Hu}}, \bibinfo {author}
  {\bibfnamefont {H.}~\bibnamefont {Xiao}}, \bibinfo {author} {\bibfnamefont
  {H.-H.}\ \bibnamefont {Wen}}, \bibinfo {author} {\bibfnamefont
  {M.}~\bibnamefont {Dzero}}, \ and\ \bibinfo {author} {\bibfnamefont {C.~C.}\
  \bibnamefont {Almasan}},\ }\href {\doibase 10.1103/PhysRevB.95.184513}
  {\bibfield  {journal} {\bibinfo  {journal} {Phys. Rev. B}\ }\textbf {\bibinfo
  {volume} {95}},\ \bibinfo {pages} {184513} (\bibinfo {year}
  {2017})}\BibitemShut {NoStop}%
\bibitem [{\citenamefont {Summers}\ \emph {et~al.}(1991)\citenamefont
  {Summers}, \citenamefont {Guinan}, \citenamefont {Miller},\ and\
  \citenamefont {Hahn}}]{SummersIEEE1991}%
  \BibitemOpen
  \bibfield  {author} {\bibinfo {author} {\bibfnamefont {L.}~\bibnamefont
  {Summers}}, \bibinfo {author} {\bibfnamefont {M.}~\bibnamefont {Guinan}},
  \bibinfo {author} {\bibfnamefont {J.}~\bibnamefont {Miller}}, \ and\ \bibinfo
  {author} {\bibfnamefont {P.}~\bibnamefont {Hahn}},\ }\href {\doibase
  10.1109/20.133608} {\bibfield  {journal} {\bibinfo  {journal} {Magnetics,
  IEEE Transactions on}\ }\textbf {\bibinfo {volume} {27}},\ \bibinfo {pages}
  {2041} (\bibinfo {year} {1991})}\BibitemShut {NoStop}%
\bibitem [{\citenamefont {Attanasio}\ \emph {et~al.}(1995)\citenamefont
  {Attanasio}, \citenamefont {Coccorese}, \citenamefont {Kushnir},
  \citenamefont {Maritato}, \citenamefont {Prischepa},\ and\ \citenamefont
  {Salvato}}]{AttanasioPhysicaC1995}%
  \BibitemOpen
  \bibfield  {author} {\bibinfo {author} {\bibfnamefont {C.}~\bibnamefont
  {Attanasio}}, \bibinfo {author} {\bibfnamefont {C.}~\bibnamefont
  {Coccorese}}, \bibinfo {author} {\bibfnamefont {V.}~\bibnamefont {Kushnir}},
  \bibinfo {author} {\bibfnamefont {L.}~\bibnamefont {Maritato}}, \bibinfo
  {author} {\bibfnamefont {S.}~\bibnamefont {Prischepa}}, \ and\ \bibinfo
  {author} {\bibfnamefont {M.}~\bibnamefont {Salvato}},\ }\href {\doibase
  http://dx.doi.org/10.1016/0921-4534(95)00616-8} {\bibfield  {journal}
  {\bibinfo  {journal} {Physica C: Superconductivity}\ }\textbf {\bibinfo
  {volume} {255}},\ \bibinfo {pages} {239 } (\bibinfo {year}
  {1995})}\BibitemShut {NoStop}%
\bibitem [{\citenamefont {Traito}\ \emph {et~al.}(2006)\citenamefont {Traito},
  \citenamefont {Peurla}, \citenamefont {Huhtinen}, \citenamefont {Stepanov},
  \citenamefont {Safonchik}, \citenamefont {Tse}, \citenamefont {Paturi},\ and\
  \citenamefont {Laiho}}]{TraitoPRB2006}%
  \BibitemOpen
  \bibfield  {author} {\bibinfo {author} {\bibfnamefont {K.}~\bibnamefont
  {Traito}}, \bibinfo {author} {\bibfnamefont {M.}~\bibnamefont {Peurla}},
  \bibinfo {author} {\bibfnamefont {H.}~\bibnamefont {Huhtinen}}, \bibinfo
  {author} {\bibfnamefont {Y.~P.}\ \bibnamefont {Stepanov}}, \bibinfo {author}
  {\bibfnamefont {M.}~\bibnamefont {Safonchik}}, \bibinfo {author}
  {\bibfnamefont {Y.~Y.}\ \bibnamefont {Tse}}, \bibinfo {author} {\bibfnamefont
  {P.}~\bibnamefont {Paturi}}, \ and\ \bibinfo {author} {\bibfnamefont
  {R.}~\bibnamefont {Laiho}},\ }\href {\doibase 10.1103/PhysRevB.73.224522}
  {\bibfield  {journal} {\bibinfo  {journal} {Phys. Rev. B}\ }\textbf {\bibinfo
  {volume} {73}},\ \bibinfo {pages} {224522} (\bibinfo {year}
  {2006})}\BibitemShut {NoStop}%
\bibitem [{\citenamefont {Hänisch}\ \emph {et~al.}(2015)\citenamefont
  {Hänisch}, \citenamefont {Iida}, \citenamefont {Kurth}, \citenamefont
  {Reich}, \citenamefont {Tarantini}, \citenamefont {Jaroszynski},
  \citenamefont {Förster}, \citenamefont {Fuchs}, \citenamefont {Hühne},
  \citenamefont {Grinenko}, \citenamefont {Schultz},\ and\ \citenamefont
  {Holzapfel}}]{JensNature2015}%
  \BibitemOpen
  \bibfield  {author} {\bibinfo {author} {\bibfnamefont {J.}~\bibnamefont
  {Hänisch}}, \bibinfo {author} {\bibfnamefont {K.}~\bibnamefont {Iida}},
  \bibinfo {author} {\bibfnamefont {F.}~\bibnamefont {Kurth}}, \bibinfo
  {author} {\bibfnamefont {E.}~\bibnamefont {Reich}}, \bibinfo {author}
  {\bibfnamefont {C.}~\bibnamefont {Tarantini}}, \bibinfo {author}
  {\bibfnamefont {J.}~\bibnamefont {Jaroszynski}}, \bibinfo {author}
  {\bibfnamefont {T.}~\bibnamefont {Förster}}, \bibinfo {author}
  {\bibfnamefont {G.}~\bibnamefont {Fuchs}}, \bibinfo {author} {\bibfnamefont
  {R.}~\bibnamefont {Hühne}}, \bibinfo {author} {\bibfnamefont
  {V.}~\bibnamefont {Grinenko}}, \bibinfo {author} {\bibfnamefont
  {L.}~\bibnamefont {Schultz}}, \ and\ \bibinfo {author} {\bibfnamefont
  {B.}~\bibnamefont {Holzapfel}},\ }\href {http://dx.doi.org/10.1038/srep17363}
  {\bibfield  {journal} {\bibinfo  {journal} {Scientific Reports}\ }\textbf
  {\bibinfo {volume} {5}},\ \bibinfo {pages} {17363} (\bibinfo {year}
  {2015})}\BibitemShut {NoStop}%
\end{thebibliography}%


\end{document}